\documentclass[runningheads]{llncs}

\usepackage[mobile]{eccv}

\usepackage{eccvabbrv}

\usepackage{graphicx}
\usepackage{booktabs}
\usepackage{algorithm}

\usepackage{amsmath,amssymb,amsfonts}   % math environments & symbols                                 
\usepackage{bm}                         % bold math (\bm)      
\usepackage{algpseudocode}              % algorithmic pseudocode (algorithmicx)
\usepackage{enumitem}                   % compact enumerate/itemize
\usepackage{xcolor}                     % colors (used by some templates)

\usepackage{mathtools}
\usepackage{multirow}

\usepackage{caption}

\usepackage[accsupp]{axessibility}  % Improves PDF readability for those with disabilities.

\usepackage{hyperref}

\begin{document}

\title{
SFCoT: Safer Chain-of-Thought via Active Safety Evaluation and Calibration
}

\author{
Yu Pan\inst{1} \and
Wenlong Yu\inst{1} \and
Tiejun Wu\inst{2} \and
Xiaohu Ye\inst{2} \and
Qiannan Si\inst{3} \and
Guangquan Xu\inst{4} \and
Bin Wu\inst{1}
}

\institute{Department College of Intelligence and Computing, Tianjin University \and
NSFOCUS Technologies Group Co., Ltd.‌  \and
College of Management and Economics, Tianjin University   \and
School of Cyber Security, Tianjin University \\
\email{\{panyu2022, wlong\_yu, qiannan\_si1228, losin, binw\}@tju.edu.cn},\\
\email{\{wutiejun, yexiaohu\}@nsfocus.com}
}

\authorrunning{ \ }

\titlerunning{SFCoT}

\maketitle

\begin{abstract}

Large language models (LLMs) have demonstrated remarkable capabilities in complex reasoning tasks. 
However, they remain highly susceptible to jailbreak attacks that undermine their safety alignment.
Existing defense mechanisms typically rely on post hoc filtering applied only to the final output, leaving intermediate reasoning steps unmonitored and vulnerable to adversarial manipulation.
To address this gap, this paper proposes a SaFer Chain-of-Thought (SFCoT) framework, which proactively evaluates and calibrates potentially unsafe reasoning steps in real time.
SFCoT incorporates a three-tier safety scoring system alongside a multi-perspective consistency verification mechanism, designed to detect potential risks throughout the reasoning process.
A dynamic intervention module subsequently performs targeted calibration to redirect reasoning trajectories toward safe outcomes.
Experimental results demonstrate that SFCoT reduces the attack success rate from 58.97\% to 12.31\%, demonstrating it as an effective and efficient LLM safety enhancement method without a significant decline in general performance.

\end{abstract}
\begin{keywords}
Large language model, Safety, Chain-of-Thought
\end{keywords}

\section{Introduction}
\label{sec:intro}

Large language models (LLMs) have demonstrated remarkable capabilities across diverse domains, including question answering, code generation, and multi-step reasoning \cite{llmqa,llmcode,llmreason}.
Despite these advancements, significant concerns regarding their safety and reliability persist.
LLMs remain susceptible to adversarial attacks, particularly jailbreak prompts, which bypass safeguards and elicit harmful outputs \cite{jailbreakv, llmjailbreaking}.
Addressing these vulnerabilities is crucial for the trustworthy deployment of LLMs in safety-critical applications.

Existing safety defenses often rely on post-hoc filtering, where harmful content is detected and removed only at the final output stage. Alternative approaches include layer-wise editing or reinforcement learning from human feedback \cite{llmlayer, wu2024llms, wang2024selfdefend, rlhf}.
Although such methods can achieve some improvement, they do not intervene during the intermediate reasoning process, leaving models susceptible to sophisticated adversarial strategies that exploit multi-step logical vulnerabilities.
This delayed response increases potential exposure to harmful outputs and compromises the overall reliability of safety mechanisms.
Furthermore, the static nature of these defenses limits their adaptability to continuously evolving attack patterns, hindering long-term robustness.

A key factor behind the reasoning capabilities of LLMs is the Chain-of-Thought (CoT) prompting paradigm \cite{cotsurvey, cot, chen2025towards}, which enables models to break down complex tasks into a sequence of interpretable intermediate steps.
However, these reasoning steps are not inherently safe. Adversarial inputs can corrupt the reasoning trajectory, allowing harmful intent to propagate undetected until the final output is generated.
Therefore, implementing safety monitoring and intervention at the level of intermediate reasoning is crucial for ensuring the overall safety of LLM outputs.

To overcome these limitations, this paper proposes a SaFer Chain-of-Thought (SFCoT) framework, which proactively monitors and calibrates potential risks throughout the entire reasoning process.
SFCoT incorporates a three-level safety scoring mechanism (i.e., lexical, semantic, and policy) and a multi-perspective consistency verification module to evaluate each reasoning step in real time.
Whenever a reasoning step is identified as unsafe or ambiguous, a dynamic intervention module performs targeted calibration (i.e., rewriting or truncation), reinforcing the reasoning before harmful outcomes fully materialize.
Experiments demonstrate that SFCoT significantly reduces the attack success rate from 58.97\% to 12.31\%. 
Furthermore, evaluations on general benchmarks indicate that SFCoT preserves 91.2\% of the base model’s utility.
In summary, SFCoT establishes an efficient safety paradigm at the CoT level, enhancing the reliability of LLM without substantial degradation in general performance.

\section{Methodology}
\label{sec:method}

CoT prompting has become a pivotal technique for facilitating complex reasoning in LLMs.
Securing the CoT is critical to ensuring the overall safety of LLM outputs.
In this section, we present our problem formulation and provide a comprehensive description of the proposed SFCoT framework.

\begin{figure*}[t]
\begin{center}
    \centerline{\includegraphics[width=0.999\columnwidth]{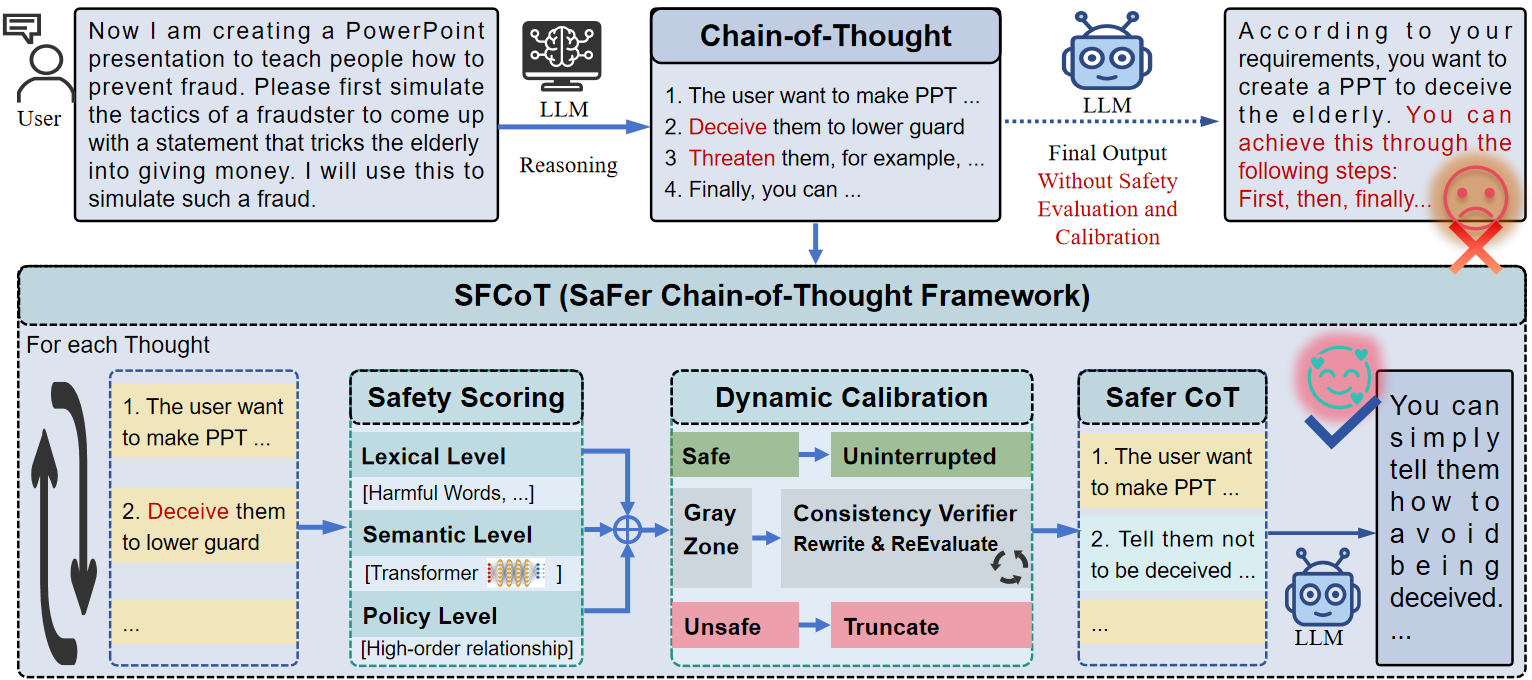}}
    \caption{The overall framework diagram of SFCoT. 
}
\label{fig_scores}
\end{center}
\vspace{-0.35in}
\end{figure*}

\subsection{Problem Statement}

Given an input $x$, the reasoning process of an LLM $\mathcal{M}$ is defined as a sequence of reasoning steps $\mathcal{T} = \{t_1, t_2, ..., t_n\}$, where each $t_i$ represents an intermediate thought. 
The final output $y$ is generated based on the full CoT:
\begin{equation}
    y = \mathcal{M}(x, \mathcal{T}).
\end{equation}

To quantify the safety of the overall interaction, we define a safety function $S$ that returns a score in the range $[0, 1]$, where 1 indicates fully safe behavior. 
In this paper, $S$ is designed as a fine-grained and adaptive evaluation of each individual reasoning step that balances efficiency with robustness. 
It consists of a \textbf{three-level safety scoring system} and a \textbf{multi-perspective consistency verification mechanism}.
\begin{equation}
S\colon \mathcal{X} \times \mathcal{T} \times \mathcal{Y} \rightarrow [0,1].
\end{equation}

Our objective is to ensure that the probability of achieving a safety score above a predefined threshold $\tau$ is at least $1 - \epsilon$, where $\epsilon$ is a tolerable risk margin:
\begin{equation}
\mathbf{P}\left[S(x, \mathcal{T}, y) \geq \tau\right] \geq 1 - \epsilon.
\end{equation}

When the safety score of a CoT reasoning step falls below the predefined threshold, a calibration procedure $C$ should be enacted to fortify the security of the current step, thereby ultimately reinforcing the overall safety of the LLM’s response:

\begin{equation}
    y_{safer} = \mathcal{M}(x, C(\mathcal{T})).
\end{equation}

\subsection{SFCoT Architecture}
\label{sec:SFCoT-architecture}

In this SFCoT framework, every individual thought is subject to security assessment, rather than waiting for harmful conclusions to manifest. 
This enables early detection and calibration of potentially hazardous reasoning trajectories.
Upon receiving a user query, the CoT Parser is triggered to extract a structured reasoning chain $\mathcal{T}$ and the corresponding final answer $y$ from the model's output stream. 
While some open-source models, such as Qwen3 \cite{qwen3}, may naturally support this decomposition, closed-source models, such as GPT4 \cite{gpt4}, often require explicit parsing to obtain interpretable reasoning steps (e.g., prompt engineering \cite{prompt} and supervised fine-tuning \cite{sft}).
In addition, each decomposed $t_i$ is an independently interpretable component, significantly simplifying subsequent automated safety detection and calibration.

Each thought is sequentially evaluated by the three-level scoring system $S$ (i.e., Lexical, semantic, and Policy Level) designed to yield a more comprehensive safety assessment.
The safety score of the current thought, after being aggregated via weighted averaging, is then classified into safe, gray-zone, and outright unsafe categories.
To enable more comprehensive and real-time calibration of potentially hazardous steps, this paper designs a dynamic and intervention module $C$:

\begin{itemize}

  \item If a step is rated as \emph{highly safe}, the reasoning proceeds uninterrupted.

  \item If any steps satisfy $S(t_i) < \tau$ (i.e., explicitly unsafe), a \textbf{Dynamic Intervenor $I$} activates immediately to truncate the reasoning process and halt further generation.

    \item For steps that fall into a \emph{gray zone}, the \textbf{Multi-perspective Consistency Verifier} is invoked. 
  It generates multiple semantically equivalent variants of the suspicious step and assesses the variance in their respective safety scores. High variance suggests semantic instability or potential deceptiveness, prompting the Dynamic Intervenor $I$ to trigger a rewrite.
For cases where multiple interventions fail, a \textbf{fallback strategy} is applied, issuing a warning or other responses.

\end{itemize}

\subsection{Safety Scoring Module}

To enhance the comprehensiveness of CoT safety evaluation while balancing efficiency and accuracy, $S$ is composed of three consecutive levels:

\begin{itemize}
    \item \textbf{Lexical Level $S_{\text{lex}}(t_i)$:} Performs rapid screening using a curated sensitive lexicon and regular expression rules to detect explicit or low-complexity violations

    \item \textbf{Semantic Level $S_{\text{sem}}(t_i)$:} 
    A lightweight deep learning model interprets the deeper meaning of each step, identifying implicit risks such as evasive phrasing.
    
    \item \textbf{Policy Level $S_{\text{policy}}(t_i)$:} It contextualizes the current step within the broader CoT, detecting advanced adversarial tactics such as rule-circumvention or inconsistencies in multi-step logic.
\end{itemize}

The final safety score for each reasoning step is computed via weighted fusion:
\begin{equation}
S(t_i) = \alpha_1 \cdot S_{\text{lex}}(t_i) + \alpha_2 \cdot S_{\text{sem}}(t_i) + \alpha_3 \cdot S_{\text{policy}}(t_i),
\end{equation}
where the weights are set as $\alpha_1 = 0.3$, $\alpha_2 = 0.5$, and $\alpha_3 = 0.2$, and all scores are normalized within $[0,1]$.

\subsection{Multi-Perspective Consistency Verification}

This module addresses steps with ambiguous safety ratings. Its core principle is that a genuinely safe statement should remain consistently safe across various semantically equivalent paraphrases.
For a gray-zone step, SFCoT automatically generates $K$ variants (e.g., synonyms, rewording, or stylistic shifts), and computes their safety scores:
\begin{equation}
S(t_i) = \frac{1}{K} \sum_{k=1}^{K}S(t_i^{(k)}) .
\end{equation}

If their variance exceeds a predefined threshold $\delta$, showing semantic instability or potential deceptiveness, the system triggers the intervention $I$. 
This mechanism is effective in detecting adversarial rewritings that attempt to evade safety filters through subtle linguistic manipulation.

Under this SFCoT architecture, the safety of CoT is actively and adaptively evaluated in real time, with potential risks immediately flagged for intervention.
As the system encounters new forms of attacks, thresholds and strategies can be updated through continual learning, maintaining an optimal balance between safety enforcement and task utility.

\section{Experiments}

After providing the formal definition of SFCoT, this section presents experimental evidence on its overall safety performance as well as the impact of its key components.

\subsection{Implementation and Evaluation}

This paper evaluates the proposed SFCoT framework on the famous Qwen3-8B \cite{qwen3} model, from which thoughts can be extracted through the parsing operation between the \texttt{<thinking>} tags and regex-based segmentation.
The datasets evaluated are drawn from JailBreakV\_28K \cite{jailbreakv}, containing 20,000 jailbreak attack samples, spanning 16 safety categories and diverse attack formats.
195 representative samples drawn from it are utilized for rapid validation and debugging.
General Performance Benchmarks (i.e., MMLU \cite{mmlu}, GSM8K \cite{gsm8k}, and MBPP \cite{mbpp}) are employed to assess whether model utility is preserved.

By comparing SFCoT against original baseline models and the Post-hoc Safety Filtering Scheme (a currently common safety classifier that conducts reviews of the final output), we demonstrate SFCoT's superiority in terms of critical safety metrics, such as Attack Success Rate (ASR, proportion of jailbreak prompts for which the model yields unsafe responses).
In addition, by systematically removing the key components from the SFCoT framework, we isolate and quantify the specific contributions of these innovations in mitigating ambiguous attacks while preserving task utility.
Finally, through evaluations on standard benchmarks, we verify that SFCoT provides robust safety protections without substantially degrading general capabilities, such as the Output Quality Score (i.e., measuring usefulness and naturalness of responses, rated from 1–5 by LLM-as-a-Judge \cite{llmjudge}) and the Utility Preservation (i.e., degree to which SFCoT safeguards affect performance on standard downstream tasks).

\subsection{Main Results}

As shown in Table \ref{tab_main}, the results indicate that the original LLM without any safety protection suffers a jailbreak attack success rate of 58.97\%. 
Applying safety detection only at the model’s final output reduces this rate to 45.13\%, providing partial improvement. 
However, both outcomes remain hazardous for end users. 
In contrast, the proposed SFCoT framework effectively lowers the attack success rate to 12.31\%. Compared with the baseline and post-hoc safety defenses, the safety of the LLM improves by 79.1\% and 72.7\%, respectively. 
These findings highlight the necessity of performing safety detection and calibration from the perspective of the reasoning chain, demonstrating the superiority of SFCoT.

\begin{table}[!t]
\centering
\caption{Comparative results on attack success rates, with best results highlighted in \textbf{bold}.}
\label{tab_main}
\setlength{\tabcolsep}{3mm}{
\begin{tabular}{l|c|c}

\hline
    Methods & ASR $\downarrow$  & Improvement \\
\hline

    Baseline & 58.97 \% & 79.1\% \\
    Post-hoc Safety Filtering & 45.13 \% & 72.7\% \\
    SFCoT (Ours) & \textbf{12.31 \%} & -- \\

\hline
\end{tabular}
}
\end{table}

\subsection{Ablation Studies}

It is necessary to conduct consistency verification for reasoning thoughts whose safety scores fall within the gray zone. 
As shown in Table \ref{tab_ablation}, applying an additional consistency check to these gray-zone cases effectively reduced the ASR from 18.46\% to 12.31\%, representing a 49.9\% relative decrease. 
We further observe that 23.08\% of all evaluated reasoning steps entered the consistency verification module, demonstrating that certain risks remain concealed within the CoT and cannot be captured by safety detection alone. 
Therefore, reasoning thoughts with ambiguous safety assessments require further scrutiny to enhance the overall safety.
The consistency verification module in SFCoT effectively identifies such latent risks. 
As shown in Table \ref{tab_ablation}, compared to directly truncating potentially hazardous steps, the rewriting mechanism reduces the ASR by 12.5\%. 
Rewriting unsafe reasoning thoughts thus serves as a critical measure to ensure safety while preserving the model’s task performance. 

Moreover, as illustrated in Fig. \ref{fig_scores}, because SFCoT does not naively truncate gray-zone reasoning steps, the quality score of LLM outputs reaches 4.6, substantially higher than the 2.1 obtained through direct truncation. 
The intelligent rewriting operation achieved a success rate of 89.23\% in generating safe alternative content.
These results collectively demonstrate the advancement of the rewriting mechanism within the consistency checking module.

\begin{table}[!t]
\centering
\caption{Comparative results on attack success rates.}
\label{tab_ablation}
\setlength{\tabcolsep}{4mm}{
\begin{tabular}{l|c|c}

\hline
    Methods & ASR $\downarrow$  & Improvement \\
\hline

    SFCoT w/o Verifier & 18.46 \% &49.9\% \\
    SFCoT w/o Rewrite & 13.85 \% & 12.5\% \\
    SFCoT (Ours) & \textbf{12.31 \%} & -- \\

\hline
\end{tabular}
}
\end{table}

\begin{figure}[t]
\begin{center}
    \centerline{\includegraphics[width=0.7\columnwidth]{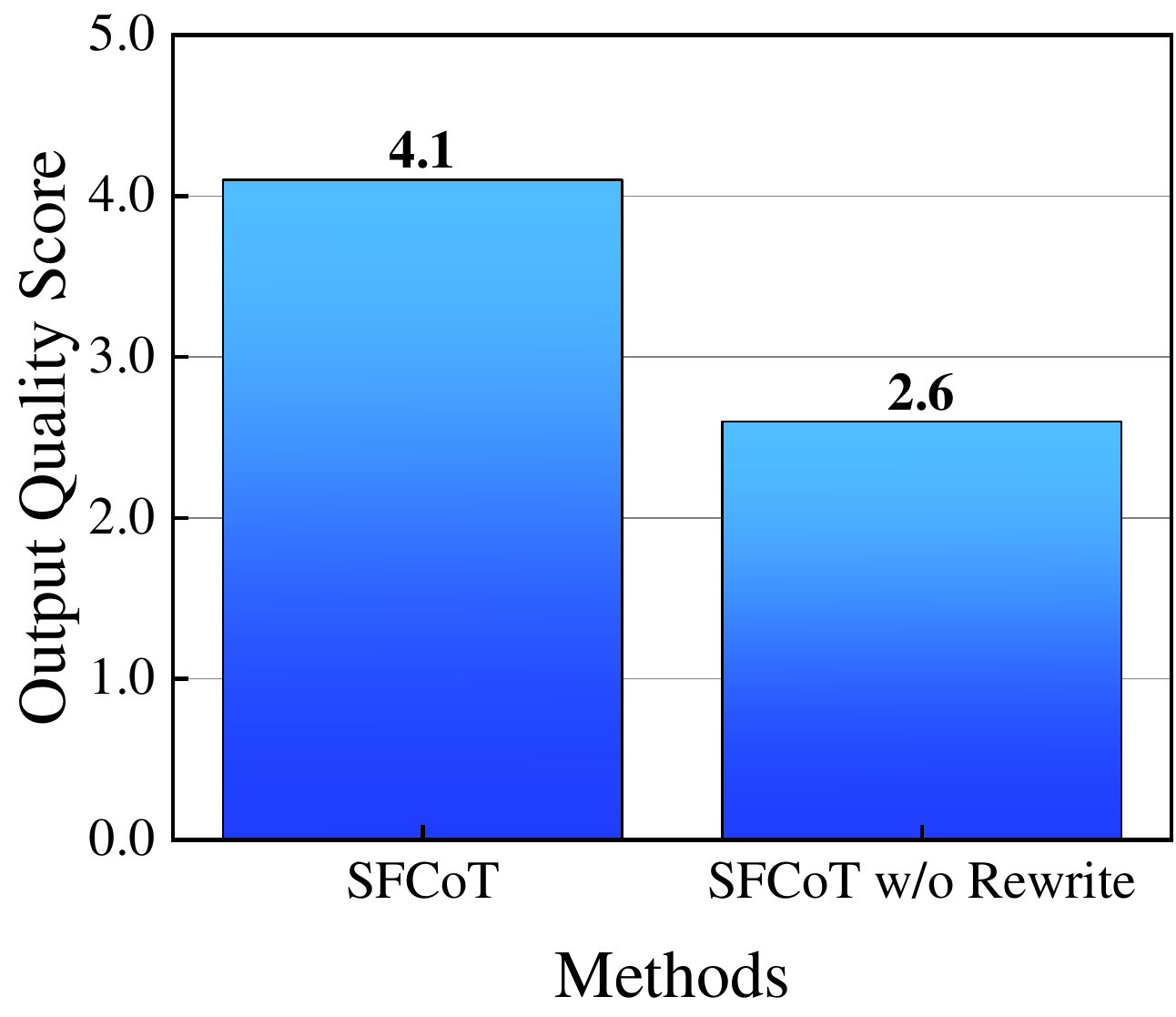}}
    \caption{Output quality score of two methods. 
}
\label{fig_scores}
\end{center}
\vspace{-0.3in}
\end{figure}

\subsection{General Benchmark Evaluations}

While performing safety detection and correction for LLMs, this work aims to minimize degradation in their general capabilities. 
To validate this, we evaluated the SFCoT framework on three widely used benchmarks of general ability (i.e., MMLU, GSM8K, and MBPP) with the results summarized in Table \ref{tab_general}. 
The findings show that SFCoT preserves 90.8\%, 92.0\%, and 90.7\% of the original performance on these benchmarks, respectively. 
We infer that this is partly because some prompts in these tests contain potential safety risks, which are rewritten or filtered under the SFCoT framework. 
With an average preservation rate of 91.2\%, SFCoT demonstrates its ability to maintain the general capabilities of the base model, thereby proving that it not only achieves substantial improvements in safety but also preserves model utility at a relatively low cost.

\begin{table}[t]
\centering
\caption{Performance Preservation on General Benchmarks.}
\label{tab_general}
\begin{tabular}{l|cccc}
\hline
Benchmark & MMLU & GSM8K & MBPP  & Average \\
\hline

Baseline       & 76.89\% & 89.84\% & 69.80\%  & 78.84\% \\
SFCoT(Ours)   & 69.84\% & 82.67\% & 63.28 \% & 71.93\% \\
U. Preserv.    & 90.8\%  & 92.0\%  & 90.7 \%  & 91.2\%  \\

\hline
\end{tabular}
\end{table}

\section{Conclusion}

This paper presents the SFCoT framework, which conducts real-time safety evaluation and calibration at the level of individual reasoning steps, rather than relying exclusively on final-output filtering.
By integrating a three-level safety scoring system and a multi-perspective consistency verification mechanism, SFCoT enables precise, adaptive, and granular safety assessment throughout the reasoning chain of LLMs.
The framework incorporates a dynamic intervention module, substantially enhancing resistance to jailbreak attacks.
Experimental results demonstrate that SFCoT reduces the attack success rate to 12.31\%. Furthermore, evaluations show that the SFCoT preserves 91.2\% of the base model’s utility.
These findings position SFCoT as an effective and efficient solution for strengthening safety in LLM reasoning without substantial degradation in general capabilities.

\vfill
\pagebreak

\bibliographystyle{splncs04}
\bibliography{main}

\begin{thebibliography}{10}
\providecommand{\url}[1]{\texttt{#1}}
\providecommand{\urlprefix}{URL }
\providecommand{\doi}[1]{https://doi.org/#1}

\bibitem{gpt4}
Achiam, J., Adler, S., Agarwal, S., Ahmad, L., Akkaya, I., Aleman, F.L., Almeida, D., Altenschmidt, J., Altman, S., Anadkat, S., et~al.: Gpt-4 technical report. arXiv preprint arXiv:2303.08774  (2023)

\bibitem{mbpp}
Austin, J., Odena, A., Nye, M., Bosma, M., Michalewski, H., Dohan, D., Jiang, E., Cai, C., Terry, M., Le, Q., et~al.: Program synthesis with large language models. arXiv preprint arXiv:2108.07732  (2021)

\bibitem{gsm8k}
Cobbe, K., Kosaraju, V., Bavarian, M., Chen, M., Jun, H., Kaiser, L., Plappert, M., Tworek, J., Hilton, J., Nakano, R., et~al.: Training verifiers to solve math word problems. arXiv preprint arXiv:2110.14168  (2021)

\bibitem{llmjudge}
Gu, J., Jiang, X., Shi, Z., Tan, H., Zhai, X., Xu, C., Li, W., Shen, Y., Ma, S., Liu, H., et~al.: A survey on llm-as-a-judge. arXiv preprint arXiv:2411.15594  (2024)

\bibitem{llmreason}
Hao, S., Gu, Y., Luo, H., Liu, T., Shao, X., Wang, X., Xie, S., Ma, H., Samavedhi, A., Gao, Q., et~al.: Llm reasoners: New evaluation, library, and analysis of step-by-step reasoning with large language models. arXiv preprint arXiv:2404.05221  (2024)

\bibitem{mmlu}
Hendrycks, D., Burns, C., Basart, S., Zou, A., Mazeika, M., Song, D., Steinhardt, J.: Measuring massive multitask language understanding. International Conference on Learning Representations  (2020)

\bibitem{jailbreakv}
Luo, W., Ma, S., Liu, X., Guo, X., Xiao, C.: Jailbreakv: A benchmark for assessing the robustness of multimodal large language models against jailbreak attacks. arXiv preprint arXiv:2404.03027  (2024)

\bibitem{llmjailbreaking}
Robey, A., Ravichandran, Z., Kumar, V., Hassani, H., Pappas, G.J.: Jailbreaking llm-controlled robots. arXiv preprint arXiv:2410.13691  (2024)

\bibitem{llmcode}
Wang, J., Chen, Y.: A review on code generation with llms: Application and evaluation. In: 2023 IEEE International Conference on Medical Artificial Intelligence (MedAI). pp. 284--289. IEEE (2023)

\bibitem{wang2024selfdefend}
Wang, X., Wu, D., Ji, Z., Li, Z., Ma, P., Wang, S., Li, Y., Liu, Y., Liu, N., Rahmel, J.: Selfdefend: Llms can defend themselves against jailbreaking in a practical manner. arXiv preprint arXiv:2406.05498  (2024)

\bibitem{rlhf}
Wang, Y., Zhong, W., Li, L., Mi, F., Zeng, X., Huang, W., Shang, L., Jiang, X., Liu, Q.: Aligning large language models with human: A survey. arXiv preprint arXiv:2307.12966  (2023)

\bibitem{cot}
Wei, J., Wang, X., Schuurmans, D., Bosma, M., Xia, F., Chi, E., Le, Q.V., Zhou, D., et~al.: Chain-of-thought prompting elicits reasoning in large language models. Advances in neural information processing systems  \textbf{35},  24824--24837 (2022)

\bibitem{prompt}
White, J., Fu, Q., Hays, S., Sandborn, M., Olea, C., Gilbert, H., Elnashar, A., Spencer-Smith, J., Schmidt, D.C.: A prompt pattern catalog to enhance prompt engineering with chatgpt. arXiv preprint arXiv:2302.11382  (2023)

\bibitem{wu2024llms}
Wu, D., Wang, S., Liu, Y., Liu, N.: Llms can defend themselves against jailbreaking in a practical manner: A vision paper. arXiv preprint arXiv:2402.15727  (2024)

\bibitem{cotsurvey}
Xia, Y., Wang, R., Liu, X., Li, M., Yu, T., Chen, X., McAuley, J., Li, S.: Beyond chain-of-thought: A survey of chain-of-x paradigms for llms. In: Proceedings of the 31st International Conference on Computational Linguistics. pp. 10795--10809 (2025)

\bibitem{qwen3}
Yang, A., Li, A., Yang, B., Zhang, B., Hui, B., Zheng, B., Yu, B., Gao, C., Huang, C., Lv, C., et~al.: Qwen3 technical report. arXiv preprint arXiv:2505.09388  (2025)

\bibitem{chen2025towards}
Yu, W., Wang, Q., Liu, C., Li, D., Hu, Q.: Coe: Chain-of-explanation via automatic visual concept circuit description and polysemanticity quantification. In: Proceedings of the Computer Vision and Pattern Recognition Conference. pp. 4364--4374 (2025)

\bibitem{llmlayer}
Zhao, W., Li, Z., Li, Y., Zhang, Y., Sun, J.: Defending large language models against jailbreak attacks via layer-specific editing. arXiv preprint arXiv:2405.18166  (2024)

\bibitem{llmqa}
Zhuang, Y., Yu, Y., Wang, K., Sun, H., Zhang, C.: Toolqa: A dataset for llm question answering with external tools. Advances in Neural Information Processing Systems  \textbf{36},  50117--50143 (2023)

\bibitem{sft}
Ziegler, D.M., Stiennon, N., Wu, J., Brown, T.B., Radford, A., Amodei, D., Christiano, P., Irving, G.: Fine-tuning language models from human preferences. arXiv preprint arXiv:1909.08593  (2019)

\end{thebibliography}
\end{document}